\documentclass{spie}
\usepackage{amsmath}
\usepackage[]{graphicx}

\begin{document}

\title{Quantum Computers and Decoherence: Exorcising the Demon from the
Machine}
\author{Daniel A. Lidar and Lian-Ao Wu
\skiplinehalf
Chemical Physics Theory Group, Chemistry Department, University of Toronto, 80~St.~George~St.,
Toronto, Ontario M5S 3H6, Canada }
\maketitle

\begin{abstract}
Decoherence is the main obstacle to the realization of quantum computers.
Until recently it was thought that quantum error correcting codes are the
only complete solution to the decoherence problem. Here we present an
alternative that is based on a combination of a decoherence-free subspace
encoding and the application of strong and fast pulses: ``encoded recoupling
and decoupling'' (ERD). This alternative has the advantage of lower encoding
overhead (as few as two physical qubits per logical qubit suffice), and
direct application to a number of promising proposals for the experimental
realization of quantum computers. 
\end{abstract}

\section{Introduction}

In the quest to construct large-scale quantum information processors, in
particular quantum computers, decoherence is still the main obstacle to
realization. Decoherence is the degradation of quantum information due to
inevitable interactions with the environment. Early skepticism \cite{Landauer:95,Unruh:95} concerning the viability of quantum computation (QC)
in the presence of decoherence was overcome by the discovery of quantum
error correcting codes (QECCs),\cite{Shor:95,Steane:96a,Kitaev:96,Laflamme:96,Preskill:97a,Knill:98,Steane:02} that flourished into a comprehensive theory that incorporates all elements
of quantum computation.\cite{Gottesman:97a} A QECC relies on
an encoding of quantum information into the state of several quantum bits
(qubits), and a closed-loop, active error diagnosis and correction procedure.\cite{Knill:97b} In principle, it is possible to correct arbitrary errors
using sufficiently large QECCs.\cite{Knill:98} In practice, however, this
may require a very large overhead, especially in terms of qubit resources.\cite{Steane:02} This is troubling in light of the substantial
difficulties associated with generating and controlling systems with very
large numbers of qubits. In addition, the theory of QECCs is rather
abstract, in that it presumes that one can execute certain logical
operations, but does not refer to the underlying Hamiltonians governing
specific physical systems. Hence it is of interest to explore alternatives
to QECCs, that are more economical in qubit resources, and that are direcly
tailored to specific quantum computer proposals.

Here we review our recent progress in developing such an alternative,
``encoded recoupling and decoupling'' (ERD),\cite{WuLidar:01,LidarWu:01,WuLidar:01a,WuLidar:01b,ByrdLidar:01a,WuLidar:02,WuByrdLidar:02,LidarWuBlais:02,WuLidar:02a,LidarWu:02} and report on some new results. ERD is based on a combination of encoding
quantum information into a decoherence-free subspace (DFS)
\cite{Zanardi:97c,Duan:98,Lidar:PRL98,Lidar:00a} (for a review see
Ref.~\citenum{LidarWhaley:03}), and the application of fast and strong
dynamical-decoupling (or ``bang-bang'', BB) pulses.\cite{Viola:98,Duan:98e,Vitali:99,Vitali:01,Zanardi:98b,Zanardi:99a,Zanardi:99d,Viola:99,Viola:99a,Viola:00a,Viola:01a,Viola:02,Agarwal:01,Search:00,ByrdLidar:01,WuLidar:01b,LidarWuBlais:02,WuByrdLidar:02,ByrdLidar:02,ByrdLidar:02a,ShiokawaLidar:02,LidarWu:02,Uchiyama:02} The DFS encoding provides a first layer of protection against
decoherence, while the BB pulses are used to efficiently reduce the
remaining decoherence. The encoded ``recoupling'' part of
ERD refers to the application of control operations that enact universal
quantum logic, in a manner that is fully compatible with the DFS encoding
and the BB\ pulses, and takes into account the experimentally available
control resources, such as the underlying system Hamiltonian \cite
{LidarWu:01}. The utility of ERD as a general method for quantum simulation,
universal QC, and decoherence suppression has also been stressed and
explored by Viola.\cite{Viola:01a} Our work builds in part on earlier
efforts to combine universal QC with DFS encoding (without BB pulses).\cite{Bacon:99a,Kempe:00,Zanardi:99a,Viola:99a}

The structure of the paper is as follows. We begin in
Section~\ref{decoherence} with a brief formal
summary of decoherence. We then give in Section~\ref{DFS} a simple
example of a DFS encoding, protecting against
collective dephasing. Section~\ref{logic} shows how to perform universal QC
on this encoding. We then briefly and formally review the dynamical
decoupling method, in Section~\ref{decoupling}. The methods are then
combined in Sections~\ref{createDFS},\ref{leakage-elim}, where we show
how BB pulses can eliminate decoherence sources beyond collective
dephasing. In
Section~\ref{offres} we present a new and somewhat surprising result: the elimination of
off-resonant transitions, that will typically be induced by BB pulses,
via BB pulses. Section~\ref{all} shows how to combine all the pieces
together, by including logic gates with the BB pulses. Concluding remarks are presented
in section \ref{conclusions}.

\section{Decoherence}
\label{decoherence}

The dynamics of an open quantum system coupled to a bath is formally
obtained from the time-ordered evolution 
\begin{equation}
U(t)=\mathcal{T}\exp (-i\int^{t}H(t^{\prime })dt^{\prime })  \label{eq:U}
\end{equation}
under the combined system-bath Hamiltonian 
\begin{eqnarray}
H &=&(H_{S}+H_{C})\otimes {I}_{B}+{I}_{S}\otimes H_{B}+H_{SB}  \nonumber
\\
H_{SB} &=&\sum_{\gamma }S_{\gamma }\otimes B_{\gamma },  \label{Htot}
\end{eqnarray}
where $I$ is the identity operator, $H_{S}$ ($H_{B}$) is the internal
Hamiltonian for the system (bath) alone, $H_{C}$ is an externally applied
control Hamiltonian, $H_{SB}$ is the system-bath interaction Hamiltonian,
and the $S_{\gamma }$ ($B_{\gamma }$) are operators acting on the system
(bath). The $S_{\gamma }$ play the role of error generators in QC. Making
the standard assumption of initially decoupled system and bath, one traces over the bath
degrees of freedom in order to obtain
the time-evolved reduced system density matrix: 
\begin{equation}
\rho (t)=\mathrm{Tr}_{B}[{U}(t)\left( \rho (0)\otimes \rho _{B}(0)\right) {U}
^{\dagger }(t)],  \label{eq:rhot}
\end{equation}
where $\rho (0)$ [$\rho _{B}(0)$] is the initial density matrix of the
system [bath]. As is well-known, $\rho (t)$ is in general a mixed state: 
\textrm{Tr}$[\rho (t)^{2}]<1$, indicating that the environment has decohered
the system (see, e.g., Ref.~\citenum{Lidar:CP01}). 

\section{Simple Example of Decoherence-Free Subspaces}
\label{DFS}

We briefly review the simplest example of a DFS. Suppose that a system of $K$
qubits (two-level systems) is coupled to a bath in a symmetric way, and
undergoes a dephasing process. This can be described by the system-bath
Hamiltonian 
\begin{equation}
  H_{SB}^{\mathrm{coll-deph}}=\sum_{j=1}^{K}\sigma^{z}_j\otimes
  B^{z}
  \label{eq:Hcol-deph}
\end{equation}
where $\sigma^{z}_j$ is the Pauli-$z$ matrix acting on the $j$th qubit and
$B^{z}$ is an arbitrary bath operator. Under this interaction qubit $j$
undergoes the transformation 
\begin{equation}
|0\rangle _{j}\rightarrow |0\rangle _{j}\qquad |1\rangle _{j}\rightarrow
e^{i\phi }|1\rangle _{j},
\end{equation}
which (after $\mathrm{Tr}_{B}$) puts a random phase $\phi $ between the
basis states $|0\rangle $ and $|1\rangle $ (eigenstates of $\sigma _{z}$
with respective eigenvalues $+1$ and $-1$). This can also be described by
the matrix $R_{z}(\phi )=\mathrm{diag}\left( 1,e^{i\phi }\right) $ acting on
the $\{|0\rangle ,|1\rangle \}$ basis. As implied by
(\ref{eq:Hcol-deph}), we assume that the phase has no space
($j$) dependence, i.e., the dephasing process is invariant under qubit
permutations. Since
the errors can be expressed in terms of the single Pauli spin matrix $\sigma
_{z}$ of the two-level system, this example is referred to as
``collective phase damping'', or ``weak collective decoherence''.\cite{Kempe:00}  The more general situation
when errors involving all three Pauli matrices are present, i.e.,
dissipation and dephasing, is referred to as ``strong
collective decoherence'' ,\cite{Kempe:00} or just
``collective decoherence''.\cite{Zanardi:97c,Lidar:PRL98,LidarWhaley:03} Without encoding a qubit
initially in an arbitrary pure state $|\psi \rangle _{j}=a|0\rangle
_{j}+b|1\rangle _{j}$ will decohere. This can be seen by calculating its
density matrix as an average over all possible values of $\phi $, 
\begin{equation}
\rho _{j}=\int_{-\infty }^{\infty }R_{z}(\phi )|\psi \rangle _{j}\langle
\psi |R_{z}^{\dagger }(\phi )\,p(\phi )d\phi ,
\end{equation}
where $p(\phi )$ is a probability density, and we assume the initial state
of all qubits to be a product state. For a Gaussian distribution, $p(\phi
)=\left( 4\pi \alpha \right) ^{-1/2}\exp (-\phi ^{2}/4\alpha )$, it is
simple to check that 
\begin{equation}
\rho _{j}=\left( 
\begin{array}{cc}
|a|^{2} & ab^{\ast }e^{-\alpha } \\ 
a^{\ast }be^{-\alpha } & |b|^{2}
\end{array}
\right) .
\end{equation}
The decay of the off-diagonal elements in the computational basis is a
signature of decoherence.

Let us now consider what happens in the two-qubit Hilbert space. The four
basis states undergo the transformation 
\begin{eqnarray}
|0\rangle _{1}\otimes |0\rangle _{2} &\rightarrow &|0\rangle _{1}\otimes
|0\rangle _{2} \\
|0\rangle _{1}\otimes |1\rangle _{2} &\rightarrow &e^{i\phi }|0\rangle
_{1}\otimes |1\rangle _{2} \\
|1\rangle _{1}\otimes |0\rangle _{2} &\rightarrow &e^{i\phi }|1\rangle
_{1}\otimes |0\rangle _{2} \\
|1\rangle _{1}\otimes |1\rangle _{2} &\rightarrow &e^{2i\phi }|1\rangle
_{1}\otimes |1\rangle _{2} .
\end{eqnarray}
Observe that the basis states $|0\rangle _{1}\otimes |1\rangle _{2}$ and $
|1\rangle _{1}\otimes |0\rangle _{2}$ acquire the same phase. This suggests
that a simple encoding trick can solve the decoherence problem. Let us define encoded
states by $|0_{L}\rangle =|0\rangle _{1}\otimes |1\rangle _{2}\equiv
|01\rangle $ and $|1_{L}\rangle =|10\rangle $. Then the state $|\psi
_{L}\rangle =a|0_{L}\rangle +b|1_{L}\rangle $ evolves under the dephasing
process as 
\begin{equation}
|\psi _{L}\rangle \rightarrow a|0\rangle _{1}\otimes e^{i\phi }|1\rangle
_{2}+be^{i\phi }|1\rangle _{1}\otimes |0\rangle _{2}=e^{i\phi }|\psi
_{L}\rangle ,
\end{equation}
and the overall phase thus acquired is clearly unimportant. This means that
the 2-dimensional subspace $DFS_{2}(0)=\mathrm{Span}\{|01\rangle ,|10\rangle
\}$ of the 4-dimensional Hilbert space of two qubits is
\emph{decoherence-free} (DF). The subspaces $DFS_{2}(2)=\mathrm{Span}\{|00\rangle \}$
and $DFS_{2}(-2)=\mathrm{Span}\{|11\rangle \}$ are also (trivially) DF,
since they each acquire a global phase as well, $1$ and $e^{2i\phi }$
respectively. Since the phases acquired by the different subspaces differ,
there is decoherence \emph{between} the subspaces.

For $K=3$ qubits a similar calculation reveals that the subspaces $
DFS_{3}(1)=\mathrm{Span}\{|001\rangle ,|010\rangle ,|100\rangle \}$ and $
DFS_{3}(-1)=\mathrm{Span}\{|011\rangle ,|101\rangle ,|110\rangle \}$ are DF,
as well the (trivial)\ subspaces $DFS_{3}(3)=\mathrm{Span}\{|000\rangle \}$
and $DFS_{3}(3)=\mathrm{Span}\{|111\rangle \}$.

More generally, let 
\begin{equation}
\lambda _{K}=\mathrm{number\ of\ }0^{\prime }\mathrm{s\ minus\ the\ number\
of\ }1^{\prime }\mathrm{s}
\end{equation}
in a computational basis state (i.e., a bitstring) over $K$ qubits. Then it
is easy to check that any subspace spanned by states with constant $\lambda
_{K}$ is DF, and can be denoted $DFS_{K}(\lambda _{K})$ in accordance with
the notation above. The dimensions of these subspaces are given by the
binomial coefficients: $d\equiv \dim [DFS_{K}(\lambda _{K})]={\binom{K}{\lambda _{K}}}$ and they each encode $\log _{2}d$ qubits.

The encoding for the ``collective phase damping'' model discussed here has
been tested experimentally. The first-ever experimental implementation of
DFSs used the $ DFS_{2}(0)$ subspace to protect against artifially induced
decoherence in a linear optics setting.\cite{Kwiat:00} The same encoding
was subsequently used to alleviate the problem of external fluctuating
magnetic fields in an ion trap quantum computing experiment \cite
{Kielpinski:01}, and figures prominently in theoretical constructions of
encoded, universal QC.\cite{LidarWu:01,LidarWu:02,ByrdLidar:01a}

\section{Encoded universal logic gates}

\label{logic}

Next we show how to perform universal QC on the encoding $DFS_{2}(0)= 
\mathrm{Span}\{|01\rangle ,|10\rangle \}$.

\subsection{Single encoded-qubit gates}

Let $X_{i},Y_{i},Z_{i}$ denote the standard Pauli matrices $\sigma
_{i}^{x},\sigma _{i}^{y},\sigma _{i}^{z}$, acting on the $i$th physical
qubit (we will use both notations interchangeably). In Ref.~\citenum{WuLidar:01} it
was shown that for the code $\{|0_{L}\rangle =|01\rangle ,|1_{L}\rangle
=|10\rangle \}$ the encoded logical operations (involving the first two
physical qubits) are

\begin{eqnarray}
\overline{X}_{12} &=&\frac{1}{2}(X_{1}X_{2}+Y_{1}Y_{2}),\quad  \nonumber \\
\overline{Y}_{12} &=&\frac{1}{2}(Y_{1}X_{2}-X_{1}Y_{2}),\quad  \nonumber \\
\overline{Z}_{12} &=&\frac{1}{2}(Z_{1}-Z_{2}).  \label{eq:bars}
\end{eqnarray}
These operations form an $su(2)$ algebra (i.e., we think of them as
Hamiltonians rather than unitary operators). We use a bar to denote logical
operations on the encoded qubits. In Refs.~\cite{WuLidar:01,LidarWu:01,WuLidar:01a}
these logical operations were denoted by $T_{12}^{\alpha }$, $\alpha \in
\{x,y,z\}$, and a detailed analysis was given on how to use typical
solid-state Hamiltonians (Heisenberg, $XXZ$, and $XY$ models, defined below) to implement
quantum logic operations using this DFS encoding. E.g., the term $
X_{1}X_{2}+Y_{1}Y_{2}$ is the spin-spin interaction in the $XY$ model, and $
Z_{1}-Z_{2}$ represents a Zeeman splitting. A static Zeeman splitting and a
controllable $XY$ interaction can be used to generate a universal set of
logic gates. Similar conclusions hold when the $XY$ interaction is replaced by a
Heisenberg \cite{LidarWu:01,Levy:01,Benjamin:01} or $XXZ$ interaction,\cite{WuLidar:01a} or even by a Heisenberg interaction that includes an
anistropic spin-orbit term.\cite{WuLidar:02} We remark that, as first shown
in Refs.~\citenum{Bacon:99a,Kempe:00}, the various types of exchange interactions can
be made universal also without any single-qubit terms (such as a Zeeman
splitting), by encoding into three or more qubits,\cite{Kempe:00,DiVincenzo:00a,Kempe:01,Kempe:01a,Vala:02} a result that has been
termed ``encoded universality''.\cite{Bacon:Sydney}

The two-qubit gate can be expressed as follows: 
\begin{eqnarray}
U_{ij}(\theta ,\phi _{i},\phi _{j}) &\equiv &\exp (i\theta X_{\phi
_{i}}X_{\phi _{j}})  \nonumber \\
&=&\cos \theta I_{i}I_{j}+i\sin \theta X_{\phi _{i}}X_{\phi _{j}},
\label{eq:Uij}
\end{eqnarray}
where 
\begin{equation}
X_{\phi }\equiv X\cos \phi +Y\sin \phi ,
\end{equation}
and $I$ is the identity operator. In the context of trapped-ion QC,\cite{Wineland:98} the
phase $\phi _{i}$ is the phase of the driving laser at the $i$th qubit,
while $\theta $ is proportional to the Rabi frequency, and can be set over a
wide range of values.\cite{Sorensen:00} Introducing the operators 
\begin{equation}
\tilde{X}_{ij}\equiv \frac{1}{2}(X_{i}X_{j}-Y_{i}Y_{j}),\quad \tilde{Y}
_{ij}\equiv \frac{1}{2}(Y_{i}X_{j}+X_{i}Y_{j})
\end{equation}
(denoted $R_{ij}^{x}$, $R_{ij}^{y}$ respectively in Refs.~\citenum{WuLidar:01,LidarWu:01,WuLidar:01a}) we can express

\begin{eqnarray}
U_{ij}(\theta ,\phi _{i},\phi _{j}) &=&\cos \theta \bar{I}+i\sin \theta
(\cos \Delta \phi _{ij}\overline{X}_{ij} +\sin \Delta \phi _{ij}\overline{Y}_{ij}+\cos \Phi _{ij}\tilde{X}
_{ij}+\sin \Phi _{ij}\tilde{Y}_{ij}),
\end{eqnarray}
where $\Phi _{ij}=$ $\phi _{i}+\phi _{j}$. It is simple to check that $ 
\tilde{X}_{ij}$ and $\tilde{Y}_{ij}$ annihilate the code subspace $
\{|0_{L}\rangle =|01\rangle ,|1_{L}\rangle =|10\rangle \}$ and have
non-trivial action (as encoded $X$ and $Y$) on the orthogonal subspace $
\{|00\rangle ,|11\rangle \}$. Therefore, upon restriction to the
DFS we can write: 
\begin{eqnarray}
U_{ij}(\theta ,\phi _{i},\phi _{j}) &\overset{\mathrm{DFS}}{\mapsto }&\bar{U}
_{ij}(\theta ,\Delta \phi _{ij})  \nonumber \\
&=&\exp (i\theta \overline{X}_{\Delta \phi _{ij}})=\cos \theta \bar{I}+i\sin
\theta \overline{X}_{\Delta \phi _{ij}}.  \nonumber \\
&&  \label{eq:Ubar}
\end{eqnarray}
The fact that $\bar{U}_{ij}$ depends only on the relative phase $\Delta \phi
_{ij}$ is crucial in the trapped-ion context: this quantity can be
controlled by adjusting the angle between the driving laser and the
interatomic axis, as well as by small adjustments of the trap voltages
(which, in turn, control the trap oscillation frequency, and hence the qubit
spacing), whereas it is much harder to control the absolute phase $\phi _{i}$,\cite{Sackett:00,Kielpinski:02} and hence also $\Phi _{ij}$. This is why
the code subspace $\{|01\rangle ,|10\rangle \}$ enjoys a preferred
status over the subspace $\{|00\rangle ,|11\rangle \}$.

The Hamiltonians in Eq.~(\ref{eq:bars}) generate the logic operations (\ref
{eq:Ubar}): 
\begin{eqnarray}
\exp (i\theta \overline{X}_{12}) &=&U_{12}(\theta ,\phi ,\phi )=\bar{U}
_{12}(\theta ,0)  \nonumber \\
\exp (i\theta \overline{Y}_{12}) &=&U_{12}(\theta ,\phi ,\phi +\frac{\pi }{2}
)=\bar{U}_{12}(\theta ,\frac{\pi }{2})  \nonumber \\
\exp (i\theta \overline{Z}_{12}) &=&\exp (i\frac{\pi }{4}\overline{Y}
_{12})\exp (i\theta \overline{X}_{12})\exp (-i\frac{\pi }{4}\overline{Y}
_{12})  \nonumber \\
&=&\bar{U}_{12}(\frac{\pi }{4},\pi /2)\bar{U}_{12}(\theta ,0)\bar{U}_{12}(- 
\frac{\pi }{4},\pi /2).  \label{eq:XYZ}
\end{eqnarray}
The third line follows from the elementary operator identity 
\begin{equation}
X_{\phi }=X\cos \phi +Y\sin \phi =e^{-i\phi Z/2}Xe^{i\phi Z/2}
\label{eq:identity}
\end{equation}
which holds for any $su(2)$ angular momentum set $\{X,Y,Z\}$, i.e.,
operators that satisfy the commutation relation $[X,Y]=2iZ$ (and cyclic
permutations thereof), in particular also the encoded operators $\{\overline{
X},\overline{Y},\overline{Z}\}$.

\subsection{Entangling gate between pairs of encoded qubits}

It is a well-known requirement of QC that in order to enact a universal set
of logic gates (that allow any unitary transformation to be implemented) it
is suffient to be able to implement all single-qubit operations (as in the
previous subsection), and to entangle pairs of qubits.\cite{Nielsen:book}
In this subsection we discuss the latter requirement.

In Ref.~\citenum{Kielpinski:02} the following unitary gate was
introduced, which is particularly suitable for QC using trapped ions: 
\begin{eqnarray}
U_{4} &=&\exp (-i\frac{\pi }{4}X_{\phi _{_{1}}}X_{\phi _{_{2}}}X_{\phi
_{_{3}}}X_{\phi _{_{4}}})  \nonumber \\
&=&\frac{1}{\sqrt{2}}\left( I_{1}I_{2}I_{3}I_{4}-iX_{\phi _{_{1}}}X_{\phi
_{_{2}}}X_{\phi _{_{3}}}X_{\phi _{_{4}}}\right)  \nonumber \\
&&\overset{\mathrm{DFS}}{\mapsto }\frac{1}{\sqrt{2}}\left( \bar{I}_{12}\bar{
I }_{34}-i\overline{X}_{\Delta \phi _{_{12}}}\overline{X}_{\Delta \phi
_{_{34}}}\right)  \nonumber \\
&=&\exp (-i\frac{\pi }{4}\overline{X}_{\Delta \phi _{_{12}}}\overline{X}
_{\Delta \phi _{_{34}}}).  \label{eq:U4}
\end{eqnarray}
This gate can be used to entangle two DFS-qubits. It involves simultaneous
control over two phase differences $\Delta \phi _{_{12}},\Delta \phi _{34}$,
and thus control over the motion of two pairs of qubits. The case $\Delta
\phi _{_{12}}=\Delta \phi _{34}=0$ was used in Ref.~\citenum{Sackett:00} to
demonstrate entanglement of four trapped-qubit qubits, but this choice is
not unique.

For general exchange Hamiltonians a different method is required. We review
the encoded recoupling method introduced in Ref.~\citenum{LidarWu:01}. The exchange
interaction quite generally has the form 
\begin{eqnarray}
H_{\mathrm{ex}} &=&\sum_{\alpha =x,y,z}\sum_{i<j}J_{ij}^{\alpha }\sigma
_{i}^{\alpha }\sigma _{j}^{\alpha } \\
&=&\sum_{i<j}J_{ij}^{-}R_{ij}^{x}+J_{ij}^{+}T_{ij}^{x}+J_{ij}^{z}\sigma
_{i}^{z}\sigma _{j}^{z},
\end{eqnarray}
where 
\begin{equation}
T_{ij}^{x}=\frac{1}{2}\left( \sigma _{i}^{x}\sigma _{j}^{x}+\sigma
_{i}^{y}\sigma _{j}^{y}\right) , \quad R_{ij}^{x}=\frac{1}{2}\left( \sigma
_{i}^{x}\sigma _{j}^{x}-\sigma _{i}^{y}\sigma _{j}^{y}\right) ,
\label{eq:TR}
\end{equation}
and $J_{ij}^{\pm }=J_{ij}^{x}\pm J_{ij}^{y}.$

The isotropic (Heisenberg) case corresponds to $J_{ij}^{\alpha }\equiv
J_{ij} $. The $XY$ model is the case $J_{ij}^{x}=J_{ij}^{y}$, $J_{ij}^{z}=0$
. The $ XXZ$ model is the case $J_{ij}^{x}=\pm J_{ij}^{y}\neq J_{ij}^{z}$. A
summary of QC proposals that fall into each category can be found in Ref.~\citenum{LidarWu:01}. The free Hamiltonian is 
\begin{equation}
H_{0}=\sum_{i}\frac{1}{2}\varepsilon _{i}\sigma _{i}^{z},
\end{equation}
where $\varepsilon _{i}$ is the single-particle spectrum. In each instance
of $H_{\mathrm{ex}}$ (Heisenberg, $XY$, $XXZ$) one typically has control
over only one type of parameter out of the set $\{J_{ij}^{\alpha
},\varepsilon _{i}\}$.

Let $J_{m}^{\alpha }\equiv J_{2m-1,2m}^{\alpha }$ ($\alpha =z,\pm $), and $
\epsilon _{m}^{\pm }\equiv \left( \varepsilon _{2m-1}\pm \varepsilon
_{2m}\right) /2$. Let $A$ and $B$ be two angular momentum operators
satisfying $su(2)$ commutation relations. Then if follows from
Eq.~(\ref{eq:identity}) that the operation of ``conjugating
by $A$'', 
\begin{eqnarray}
C_{A}\circ \exp (iB) &\equiv &\exp (-iA\pi /2)\exp (iB)\exp (iA\pi /2) 
\nonumber \\
&=&\exp (-iB)
\end{eqnarray}
causes $B$'s sign to be flipped.

Consider now an $XXZ$-type Hamiltonian where the $J_{m}^{+}$ parameters are
controllable but $\epsilon _{m}^{-}$ and $J_{m}^{z}$ are fixed. For
simplicity let us consider just the case $J_{m}^{-}=0$. Then we can rewrite 
\begin{equation}
H=H_{0}+H_{\mathrm{ex}}=\sum_{m=1}^{N/2}\epsilon
_{m}^{-}T_{m}^{z}-J_{m}^{z}T_{m}^{z}T_{m+1}^{z}+J_{m}^{+}T_{m}^{x},
\end{equation}
where we have omitted a constant term. The important point is now that $
T_{m}^{x}$ and $T_{m}^{z}$ satisfy $su(2)$ commutation relations. Therefore by using recoupling through
``conjugation by $T_{m}^{x}$'' we can selectively turn on and off the
single-encoded-qubit rotation $T_{m}^{z}$ and the encoded-Ising interaction $
T_{m}^{z}T_{m+1}^{z}$. This example of ``encoded selective recoupling''
establishes that encoded universal computation in the $XXZ$ model can be
done using control over the $J_{m}^{+}$ parameters alone.

Next, consider the $XY$ model, with controllable $J_{ij}^{+}$, but fixed $
\varepsilon _{i}$. To implement encoded single-qubit operations, we can use
the same encoded recoupling method as for the $XXZ$ model. As for encoded
two-qubit operations, we now no longer have the $\sigma _{i}^{z}\sigma
_{j}^{z}$ terms. Since the $XY$ model with nearest-neighbor interactions can
be shown not to be universal,\cite{WuLidar:01a} we turn on also next-nearest
neighbor $J_{ij}^{+}$ terms (these can still be nearest-neighbor in a 2D
hexagonal geometry). First note that $C_{T_{12}^{x}}\circ T_{23}^{x}=i\sigma
_{1}^{z}\sigma _{2}^{z}T_{13}^{x}$. Now assume we can control $J_{13}^{+}$;
then, using conjugation by $\pi /4$: $C_{\frac{1}{2}T_{13}^{x}}\circ \left(
C_{T_{12}^{x}}\circ T_{23}^{x}\right) =\sigma _{2}^{z}(\sigma
_{3}^{z}-\sigma _{1}^{z})/2$. Since $\sigma _{1}^{z}\sigma _{2}^{z}$ is
constant on the code subspace it can be ignored. On the other hand, $\sigma
_{2}^{z}\sigma _{3}^{z}$ again acts as $-T_{1}^{z}T_{2}^{z}$, i.e., as an
encoded $\sigma ^{z}\otimes \sigma ^{z}$. This establishes universal encoded
computation in the $XY$ model.

Taken together, the results in this section show how universal QC can be
implemented using qubits encoded into a DFS offering protection against
collective dephasing, while using only reasonable models of physically
controllable Hamiltonians. We now move on to a discussion of how to reduce
additional sources of decoherence.

\section{Dynamical decoupling pulses}

\label{decoupling}

We briefly review the dynamical decoupling method, introduced in
Ref.~\citenum{Viola:98}, and further developed in Refs.~\citenum{Duan:98e,Vitali:99,Vitali:01,Zanardi:98b,Zanardi:99a,Zanardi:99d,Viola:99,Viola:99a,Viola:00a,Viola:01a,Viola:02,Agarwal:01,Search:00,ByrdLidar:01,WuLidar:01b,LidarWuBlais:02,WuByrdLidar:02,ByrdLidar:02,ByrdLidar:02a,ShiokawaLidar:02,LidarWu:02,Uchiyama:02}.
A set of ``symmetrization'' operations is
chosen such that they form a discrete subgroup of the full unitary group of
operations on the Hilbert space of the system. Denote this group $\mathcal{G}
=\{V_{j}\}_{j=1}^{|\mathcal{G}|}$ where $V_{1}=I$. The cycle time is $T_{c}=|
\mathcal{G}|\Delta t$, where $|\mathcal{G}|$ is the number of symmetrization
operations, and $\Delta t$ is the time that the system evolves freely
between operations under $U_{0}(t)\equiv \exp (-iH_{S}t)$. The symmetrized
evolution after a single cycle is given by 
\begin{equation}
U(T_{c})=\prod_{j=1}^{|\mathcal{G}|}V_{j}^{\dagger }U_{0}(\Delta
t)V_{j}\equiv e^{iH_{\mathrm{eff}}T_{c}},  \nonumber  \label{exactevol}
\end{equation}
where the evolution under $H_{SB}+H_{B}$ has been neglected during
application of the pulses $V_{j}$, which quantifies the sense in which these
pulses have to be fast and strong (for details see, e.g, Refs.~\citenum{Viola:98,Vitali:01,ShiokawaLidar:02,LidarWu:02,Uchiyama:02}). $H_{\mathrm{
eff}}$ denotes the resulting effective Hamiltonian, that can be computed to
any order using the Magnus expansion.\cite{Ernst:book} In the BB limit one
is interested in the evolution $U(NT_{c})$ after $N$ cycles, such that $
N\rightarrow \infty $ and $\Delta t\rightarrow 0$ while $N\Delta t$ is
finite. As shown in Ref.~\citenum{Viola:99} one can then retain only the first order
Magnus term, approximating $H_{\mathrm{eff}}$ by 
\begin{equation}
H_{\mathrm{eff}}=\frac{1}{|\mathcal{G}|}\sum_{j=1}^{|\mathcal{G}
|}V_{j}^{\dagger }HV_{j}\equiv \Pi _{\mathcal{G}}(H).  \label{eq:Heff}
\end{equation}
The map $\Pi _{\mathcal{G}}$ is the projector into the centralizer $Z$ of $
\mathcal{G}$, defined as 
\begin{equation}
Z(\mathcal{G})=\{X|\;V_{j}^{\dagger }XV_{j}=X,\;\forall V_{j}\in \mathcal{G}
\}.  \nonumber
\end{equation}
Since the pulses $g_{j}$ are unitary the centralizer equals the commutant in
this case: 
\begin{equation}
Z(\mathcal{G})=\{X|\;[X,V_{j}]=0,\;\forall V_{j}\in \mathcal{G}\}.  \nonumber
\end{equation}
It is clear that $\Pi _{\mathcal{G}}$ commutes with the adjoint action $
V_{j}\cdot V_{j}^{\dagger }$ for all $j$, so that if $\mathcal{G}$ is
generated by $\{I,H_{S},S_{\gamma }\}$, the evolution will proceed without
the operators $S_{\gamma }$ affecting the system as errors: the evolution
has been \emph{symmetrized} with respect to $\mathcal{G}$.\cite{Zanardi:98b} For a geometric interpretation of this symmetrization see Ref.~\citenum{ByrdLidar:01}, where it was also pointed out that while the BB pulses
should have the effect of rotating $H$ to the vertices of a symmetric
object, they need in fact not form a group.

\section{Creating collective dephasing conditions using decoupling pulses:
reducing decoherence during storage}

\label{createDFS}

One of the important advantages of the DFS encoding $\{|01\rangle
,|10\rangle \}$ is that it is immune to collective dephasing. However,
other sources of decoherence inevitably remain. In this and the following
section, we algebraically classify all additional decoherence effects and
show how they can be eliminated.

\subsection{Creating collective dephasing on a pair of qubits}

First, let us analyze the effect of breaking the collective dephasing
symmetry, by considering a system-bath interaction of the form 
\begin{equation}
H_{SB}^{\mathrm{deph}(2)}=Z_{1}\otimes B_{1}^{z}+Z_{2}\otimes B_{2}^{z}
\end{equation}
where $B_{1}^{z},B_{2}^{z}$ are arbitrary bath operators [compare to Eq.~(\ref{eq:Hcol-deph})]. This describes a
general dephasing interaction on two qubits. The source of such dephasing
during storage can be long wavelength, randomly fluctuating ambient magnetic
fields,\cite{Kielpinski:01} that randomly shift the relative phase between
the qubit $|0\rangle $ and $|1\rangle $ states through the Zeeman effect.
The interaction can be rewritten as a sum over a collective dephasing term $
Z_{1}+Z_{2}$ and another, differential dephasing term $Z_{1}-Z_{2}$, that is
responsible for errors on the DFS: 
\begin{equation}
H_{SB}^{\mathrm{deph}(2)}=\left( Z_{1}+Z_{2}\right) \otimes B_{\mathrm{col}
}^{z}+\left( Z_{1}-Z_{2}\right) \otimes B_{\mathrm{dif}}^{z}.
\end{equation}
Here $B_{\mathrm{col}}^{z}=\left( B_{1}^{z}+B_{2}^{z}\right) /2$ and $B_{
\mathrm{dif}}^{z}=\left( B_{1}^{z}-B_{2}^{z}\right) /2$. If $B_{\mathrm{dif}
}^{z}$ were zero then there would only be collective dephasing and the DFS
encoding would offer perfect protection. However, in general $B_{\mathrm{dif}
}^{z}\neq 0$, and the DFS encoding will not suffice to offer complete
protection.

The crucial observation is that, since $Z_{1}-Z_{2}\propto \overline{Z}_{12}$
[recall Eq.~(\ref{eq:bars})], the offending term causes \emph{logical}
errors on the DFS.\cite{ByrdLidar:01a} As shown in Refs.~\citenum{WuLidar:01b,Viola:01a}, then the problem of $B_{\mathrm{dif}}^{z}\neq 0$
can be solved using a series of pulses that symmetrize $H_{SB}^{\mathrm{deph}
(2)}$ such that only the collective term remains. To do so note that since
the offending term $\propto \overline{Z}_{12}$, it anticommutes with $
\overline{X}_{12}=\frac{1}{2}(X_{1}X_{2}+Y_{1}Y_{2})$. At the same time $
\overline{X}_{12}$ commutes with $Z_{1}+Z_{2}$. This allows us to flip the
sign of the offending term by using a pair of $\pm \pi /2$ pulses in $
\overline{X}_{12}$, while leaving only the collective term. Evolution with
the flipped sign followed by unaltered evolution leads to cancellation of
the offending term. Specifically:\cite{WuLidar:01b}

\begin{equation}
e^{-iH_{SB}\tau }e^{-i\frac{\pi }{2}\overline{X}_{12}}e^{-iH_{SB}\tau }e^{i
\frac{\pi }{2}\overline{X}_{12}}=e^{-i(Z_{1}+Z_{2})\otimes B_{\mathrm{col}
}^{z}2\tau },
\end{equation}
or, in terms of gates: 
\begin{equation}
e^{-iH_{SB}\tau }\bar{U}_{12}(-\frac{\pi }{2},0)e^{-iH_{SB}\tau }\bar{U}
_{12}(\frac{\pi }{2},0)=e^{-i(Z_{1}+Z_{2})\otimes B_{\mathrm{col}}^{z}2\tau
},  \label{eq:sym}
\end{equation}
where $\bar{U}_{ij}(\theta ,\Delta \phi _{ij})$ was defined in Eq.~(\ref
{eq:Ubar}). This equation means that the system-bath coupling effectively
looks like collective dephasing at the end of the pulse sequence. Thus, the
system is periodically (every $2\tau $) projected into the DFS.

In order for the the procedure described in Eq.~(\ref{eq:sym}) to work, the
gate $\bar{U}_{12}(\pm \frac{\pi }{2},0)$ must be executed at a timescale
faster than the cutoff frequency associated with the fluctuating (magnetic)
fields causing the differential dephasing term in $H_{SB}^{\mathrm{\ deph}
(2)}$.

\subsection{Creating collective dephasing on a block of four qubits}

So far we have discussed creation of collective dephasing conditions on a
single DFS\ qubit. However, it is essential for the reliable execution of an
entangling logic gate to have collective dephasing over all four qubits
participating in the gate, even if only two are coupled at a time. A
procedure for creating collective \emph{decoherence} conditions over
blocks of $3,4,6$ and $8$ qubits was given in Ref.~\citenum{WuLidar:01b}. Here we
show how to do the same for a block of 4 qubits with collective dephasing.

Let us start with a general dephasing Hamiltonian on $N$ qubits, and rewrite
it in terms of nearest-neighbor sums and differences: 
\begin{eqnarray}
H_{SB}^{\mathrm{deph}} &=&\sum_{i=1}^{N}Z_{i}\otimes B_{i} \\
&=&\sum_{j=1}^{N/2}\left( Z_{2j}+Z_{2j-1}\right) \otimes B_{2j}^{+}+\left(
Z_{2j}-Z_{2j-1}\right) \otimes B_{2j}^{-},
\end{eqnarray}
where $B_{2j}^{\pm }\equiv (B_{2j}\pm B_{2j-1})/2$. As noted above, $
Z_{2j}-Z_{2j-1}\propto \overline{Z}_{2j-1,2j}$, so that to eliminate all
nearest-neighbor differences of the form $\left( Z_{2j}-Z_{2j-1}\right) $ we
can use the collective decoupling pulse $X_{nn}=\bigotimes_{j=1}^{N/2}e^{i 
\frac{\pi }{2}\overline{X}_{2j-1,2j}}$:

\begin{equation}
e^{-iH_{SB}\tau }X_{nn}e^{-iH_{SB}\tau }X_{nn}^{\dagger }=e^{-i2\tau
\sum_{j=1}^{N/2}\left( Z_{2j}+Z_{2j-1}\right) \otimes B_{2j}^{+}},
\end{equation}
or, in gate terms:

\[
e^{-iH_{SB}\tau }\left[ \bigotimes_{j=1}^{N/2}\bar{U}_{2j-1,2j}(-\frac{\pi }{
2},0)\right] e^{-iH_{SB}\tau }\left[ \bigotimes_{j=1}^{N/2}\bar{U}_{2j-1,2j}(
\frac{\pi }{2},0)\right] =e^{-i2\tau \sum_{j=1}^{N/2}\left(
Z_{2j}+Z_{2j-1}\right) \otimes B_{2j}^{+}}.
\]
The next step is to eliminate next-nearest neighbor differential terms. To
this end let us rewrite the outcome of the $X_{nn}$ pulse in terms of sums
and differences over blocks of four ions: 
\begin{eqnarray}
\sum_{j=1}^{N/2}\left( Z_{2j}+Z_{2j-1}\right) \otimes B_{2j}^{+}
&=&\sum_{j=1}^{N/2}\left[ Z_{2j+2}+Z_{2j+1}+Z_{2j}+Z_{2j-1}\right] \otimes
B_{2j}^{+,+}  \nonumber \\
&+&\sum_{j=1}^{N/2}\left[ (Z_{2j+2}-Z_{2j})+(Z_{2j+1}-Z_{2j-1})\right]
\otimes B_{2j}^{+,-},
\end{eqnarray}
where $B_{2j}^{+,\pm }\equiv (B_{2j+2}^{+}\pm B_{2j}^{+})/2$. The term in
the first line contains only the desired block-collective dephasing over $4$
ions. The term in the second line contains undesired differential dephasing
terms that we wish to eliminate. But these terms once again have the
appearance of encoded $Z$ operators, between next-nearest neighbor ion
pairs. Therefore we need to apply a second collective pulse $
X_{nnn}=\bigotimes_{j=1}^{N/2}e^{i\frac{\pi }{2}\overline{X}_{2j-1,2j+1}}e^{i
\frac{\pi }{2}\overline{X}_{2j,2j+2}}$, that applies encoded $X$ operators
on these qubit pairs. At this point we are left just with collective
dephasing terms on blocks of $4$ qubits, as required: 
\begin{eqnarray}
e^{-i2\tau \sum_{j=1}^{N/2}\left( Z_{2j}+Z_{2j-1}\right) \otimes B_{2j}^{+}}
\left[ \bigotimes_{j=1}^{N/2}\bar{U}_{2j-1,2j+1}(-\frac{\pi }{2},0)\bar{U}
_{2j,2j+2}(-\frac{\pi }{2},0)\right]  &\times &  \nonumber
\label{eq:create4} \\
e^{-i2\tau \sum_{j=1}^{N/2}\left( Z_{2j}+Z_{2j-1}\right) \otimes B_{2j}^{+}}
\left[ \bigotimes_{j=1}^{N/2}\bar{U}_{2j-1,2j+1}(\frac{\pi }{2},0)\bar{U}
_{2j,2j+2}(\frac{\pi }{2},0)\right]  &=&e^{-i4\tau \sum_{j=1}^{N/2}\left(
Z_{2j+2}+Z_{2j+1}+Z_{2j}+Z_{2j-1}\right) \otimes B_{2j}^{+,+}}.  \nonumber \\
&&
\end{eqnarray}
This pulse sequence is important to ensure that collective dephasing
conditions will prevail during the execution of logic gates between DFS\
qubits.

\section{Reduction of all remaining decoherence on a single DFS\ qubit
during logic gate execution}

\label{leakage-elim}

The reduction of differential dephasing errors, as in the previous
subsection, is particularly relevant for storage errors. However, this is
only the first step. Additional sources of decoherence may take place during
storage, and in particular during the execution of logic gates. It is useful
to provide a complete algebraic classification of the possible decoherence
processes. This will allow us to see what can be done using decoupling
pulses. To this end let us now write the system-bath Hamiltonian on two
physical qubits in the general form 
\begin{equation}
H_{SB}=H_{\mathrm{Leak}}+H_{\mathrm{Logi}}+H_{\mathrm{DFS}}
\end{equation}
where 
\begin{eqnarray}
H_{\mathrm{DFS}} &=&\mathrm{Span}\{\frac{ZI+IZ}{2},\frac{XY+YX}{2},\frac{
XX-YY}{2},ZZ,II\}  \nonumber \\
H_{\mathrm{Leak}} &=&\mathrm{Span}\{XI,IX,YI,IY,XZ,ZX,YZ,ZY\}  \nonumber \\
H_{\mathrm{Logi}} &=&\mathrm{Span}\{\bar{X}=\frac{XX+YY}{2},\bar{Y}=\frac{
YX-XY}{2},\bar{Z}=\frac{ZI-IZ}{2}\}  \label{eq:class}
\end{eqnarray}
where $I$ is the identity operator, $XZ\equiv X_{1}Z_{2}$ (etc.), and where $
\mathrm{Span}$ means a linear combination of these operators tensored with
bath operators. The $16$ operators in Eq.~(\ref{eq:class}) form a complete
basis for all $2$-qubit operators. This classification, first introduced in 
Ref.~\citenum{ByrdLidar:01a}, has the following significance. The operators in $H_{
\mathrm{DFS}}$ either vanish on the DFS, or are proportional to identity on
it. In either case their effect is to generate an overall phase on the DFS,
so they can be safely ignored from now on. The operators in $H_{\mathrm{Leak}
}$ are the \emph{leakage errors}: terms that cause transitions between
states inside and outside of the DFS. A universal and efficient decoupling
method for eliminating such errors, for arbitrary numbers of (encoded)\
qubits was given in Ref.~\citenum{WuByrdLidar:02}. Finally, the operators in $H_{
\mathrm{Logi}}$ have the form of logic gates on the DFS. However, these are
undesired logic operations, since they are coupled to the bath, and thus
cause decoherence.

In the previous subsection we showed how to eliminate the logical error $
\bar{Z}$, but we see now that this was only one error in a much larger set.
To deal with the additional errors it is useful at this point to introduce a
more compact notation for the pulse sequences. We denote by $[\tau ]$ a
period of evolution under the free Hamiltonian, i.e., $U(\tau )\equiv \exp
(-iH_{SB}\tau )\equiv \lbrack \tau ]$, and further denote 
\begin{equation}
P\equiv \bar{U}_{12}(-\frac{\pi }{2},0)=\exp (-i\frac{\pi }{2}\overline{X}
_{12}).
\end{equation}
Thus Eq.~(\ref{eq:sym}) can be written as: 
\begin{equation}
\exp [-i(B_{1}^{z}+B_{2}^{z})(Z_{1}+Z_{2})\tau ]=[\tau ,P,\tau ,P^{\dagger
}].
\end{equation}
As a first step in dealing with the additional errors, note that the
symmetrization procedure $[\tau ,P,\tau ,P^{\dagger }]$ can in fact achieve
more than just the elimination of the differential dephasing $Z_{1}-Z_{2}$
term. Since $\overline{X}_{12}$ also anticommutes with $\overline{Y}_{12}=
\frac{1}{2}(Y_{1}X_{2}-X_{1}Y_{2})\in H_{\mathrm{Logi}}$, if such a term
appears in the system-bath interaction it too will be eliminated using the
same procedure.

So far we have used a $\frac{\pi }{2}\overline{X}_{12}$ pulse.
Interestingly, the Hamiltonian $\overline{X}_{12}$ can also be used to
eliminate all leakage errors.\cite{ByrdLidar:01a} To see this, note that $
\bar{U}_{12}(\pm \pi ,0)=\exp (\pm i\pi \overline{X}_{12})=Z_{1}Z_{2}$. This
operator anticommutes with \emph{all} terms in $H_{\mathrm{Leak}}$. Hence it
too can be used in a parity-kick pulse sequence, that will eliminate all the
leakage errors.

At this point we are left with just a single error:\ $\overline{X}
_{12}\otimes B$ itself, in $H_{\mathrm{Logi}}$. Clearly, we cannot use a
pulse generated by $\overline{X}_{12}$ to eliminate this error. Instead, to
deal with this error we need to introduce one more pulse pair that
anticommutes with $\overline{X}_{12}$, e.g., $\exp (\pm i\frac{\pi }{2} 
\overline{Y}_{12})=\bar{U}_{12}(\pm \frac{\pi }{2},\frac{\pi }{2})$.

Let us now see how to combine all the decoherence elimination pulses into
one efficient sequence. First we introduce the abbreviations 
\begin{eqnarray}
\Pi  &\equiv &\bar{U}_{12}(\pm \pi ,0)=\exp (\pm i\pi \overline{X}_{12})=\Pi
^{\dagger }=PP  \nonumber \\
Q &\equiv &\bar{U}_{12}(-\frac{\pi }{2},\frac{\pi }{2})=\exp (-i\frac{\pi }{2
}\overline{Y}_{12})  \nonumber \\
\Lambda  &\equiv &\bar{U}_{12}(\pm \pi ,\frac{\pi }{2})=\exp (\pm i\pi 
\overline{Y}_{12})=\Lambda ^{\dagger }=QQ  \label{eq:PQetc}
\end{eqnarray}
As argued above, the $\pi $ pulse $\Pi $ eliminates $H_{\mathrm{Leak}}$: 
\begin{equation}
\exp [-i(H_{\mathrm{Logi}}+H_{\mathrm{DFS}})2\tau ]=[\tau ,\Pi ,\tau ,\Pi ].
\end{equation}
Now let us discuss adding the extra pulses needed to achieve full
decoherence elimination. The $\pi /2$ pulse $P$ eliminates $\bar{Y}$ and $
\bar{Z}$ in $H_{\mathrm{Logi}}$. Combining this with the sequence for
leakage elimination we have the sequence of $4$ pulses: 
\begin{eqnarray}
e^{-i(H_{\mathrm{DFS}}+\bar{X}\otimes B_{\bar{X}})4\tau } &=&[U(\tau )\Pi
U(\tau )\Pi ]P^{\dagger }[U(\tau )\Pi U(\tau )\Pi ]P  \nonumber \\
&=&[\tau ,\Pi ,\tau ,P,\tau ,\Pi ,\tau ,P^{\dagger }],  \label{eq:4pulses}
\end{eqnarray}
(where we have used $\Pi P^{\dagger }=P$, $\Pi P=P^{\dagger }$).

If we wish to entirely eliminate decoherence then we are left just with
getting rid of the logical error due to $\overline{X}$. To eliminate it we
now combine with the $\bar{Y}$-direction, $\pi /2$ pulse, $Q$:

\begin{eqnarray}
e^{-iH_{\mathrm{DFS}}8\tau } &=&[U(\tau )\Pi U(\tau )PU(\tau )\Pi U(\tau
)P^{\dagger }]Q^{\dagger } [U(\tau )\Pi U(\tau )PU(\tau )\Pi U(\tau
)P^{\dagger }]Q  \nonumber \\
&=&[\tau ,\Pi ,\tau ,P,\tau ,\Pi ,\tau ,P^{\dagger },Q^{\dagger }, \tau ,\Pi
,\tau ,P,\tau ,\Pi ,\tau ,P^{\dagger },Q]
\end{eqnarray}
\label{eq:10pul}
which takes ten pulses. Unfortunately it is not possible to compress this
further, since $P^{\dagger }Q=(i\overline{X})(-i\overline{Y})=i\overline{Z}$
and $P^{\dagger }Q^{\dagger }=-i\overline{Z}$, neither of which cannot be
generated directly (in one step) from the available gate $\bar{U}
_{ij}(\theta ,\Delta \phi _{ij})=\cos \theta \bar{I}+i\sin \theta \overline{X
}_{\Delta \phi _{ij}}$.

\section{Elimination of Off-Resonant Transitions}
\label{offres}

One important caveat in our discussion so far is that, because we need very strong and fast pulses,
our gate operation may become imperfect. Specifically, off-resonant coupling
may become important. This can cause \emph{unitary} leakage errors from the
DFS. These can in turn be reduced using the methods in
Refs.~\citenum{Palao:02,Tian:00}. Here we present an alternative and
new method, that uses BB pulses generated in terms of the system Hamiltonian.

Consider an $N$-level system Hamiltonian  

\[
H_{0}=\sum_{i=1}^{N}E_{i}\left| i\right\rangle \left\langle i\right| 
\]
where the levels $i=1,2$ denote our qubit. If all parameters are
fixed, the evolution $U_{0}(t)=\exp(-iH_{0}t)$ is
always on. If we turn on an interaction $H_{I},$ then the total
Hamiltonian is $H=H_{0}+H_{I},$ which can be written as

\[
H=\sum_{i=1}^{N}H_{ii}\left| i\right\rangle \left\langle i\right|
+\sum_{i>j>2}^{N}(H_{ij}\left| i\right\rangle \left\langle j\right|
+h.c.)+H_{L}
\]
where the leakage term is 
\[
H_{L}=\left| 1\right\rangle \sum_{i=3}^{N}H_{1i}\left\langle i\right|
+\left| 2\right\rangle \sum_{i=3}^{N}H_{2i}\left\langle i\right| +h.c.
\]
Now note that
\[
U_0^{\dagger }(t)HU_0(t)=\sum_{i=1}^{N}H_{ii}\left| i\right\rangle \left\langle
i\right| +\sum_{i>j=1}^{N}(e^{-i(E_{i}-E_{j})t}H_{ij}\left| i\right\rangle
\left\langle j\right| + h.c.) .
\]
Using this we now show how to eliminate each of
the leakage terms one by one. First, we eliminate $\left| 1\right\rangle
H_{13}\left\langle 3\right| +h.c.$ by the BB sequence
\[
2H^{\prime }=H+U_{0}^{\dagger }(\frac{\pi }{E_{1}-E_{3}})HU_{0}(\frac{\pi }{%
E_{1}-E_{3}})
\]
The new Hamiltonian $H^{\prime }$ does not contain  $\left| 1\right\rangle
H_{13}\left\langle 3\right| +h.c.$ Next, we eliminate $\left| 2\right\rangle
H_{23}^{\prime }\left\langle 3\right| +h.c.$ by $U_{0}(\frac{\pi }{E_{2}-E_{3}
}),$where $H_{23}^{\prime }=\exp (-i\frac{\pi (E_{2}-E_{3})}{E_{1}-E_{3}}
)H_{23}$. We can clearly repeat this procedure so as to eliminate all
leakage. \emph{The crucial point is that since we have used the system
  Hamiltonian to generate BB pulses, off-resonant transitions do not
  take place}: $H_0$ has no matrix elements between different levels.

A concern is how we to obtain $U_{0}^{\dagger }(t)=\exp(+iH_{0}t)$. This problem is shared by other methods dealing
with the same problem.\cite{Tian:00} In principle it can be solved
provided the level spacings are rationally related. Note in this
context that usually we do not need to eliminate all $H_{1i}$, since $H_{1i}\gg
H_{1i+1}$ and typically decrease exponentially according to time-independent
perturbation theory. We have assumed that there is no degeneracy. If
there is, then all degenerate transitions will be eliminated
simultaneously, so the procedure is simplified. 

\section{Combining logic gates with decoupling pulses}

\label{all}

So far we have discussed computation using the encoded recoupling method
(Section \ref{logic}), and encoded decoupling (Sections \ref{createDFS},\ref
{leakage-elim}). We now put the two together in order to obtain the full
ERD\ picture. At least two methods are available for combining quantum
computing operations with the sequences of decoupling pulses we have
presented above. For a general analysis of this issue see Ref.~\citenum{Viola:99a}.

\subsection{Fast + Strong Gates Method}

The decoupling pulse sequences given in Sec.~\ref{leakage-elim}
``stroboscopically'' create collective
dephasing conditions at the conclusion of each cycle. As noted above, this
is equivalent to a periodic projection into the DFS. This property allows
for ``stroboscopic'' quantum computation at
the corresponding projection times.\cite{Viola:99a} Here the computation
pulses need to be synchronized with the decoupling pulses, and inserted at
the end of each cycle. Because of the conditions on the validity of
the BB method,\cite{Viola:98} the amount of time available for implementation of a
logic gate is no more than the bath correlation time $\tau _{c}=2\pi /\omega
_{c}$. (An exception to this rule is the case of $1/f$
  noise.\cite{ShiokawaLidar:02}) Assuming the dominant decoherence contributions not accounted for by
the DFS encoding to come from differential dephasing (setting the $\tau _{c}$
time-scale), and given that we already assumed that we can use pulses with
interval $\Delta t\ll $ $\tau _{c}$, it is consistent to assume that we can
then also perform logic gates on the same time scale.

\subsection{Fast + Weak Gates Method}

There may be an advantage to using fast but weak pulses for the logic gates,
while preserving the fast + strong property of the decoupling pulses. To see
how to combine logic gates with decoupling in this case, let us denote by $
H_{S}=X_{\phi _{i}}X_{\phi _{j}}$ the controllable system Hamiltonian that
generates the entangling gate $U_{ij}(\theta ,\phi _{i},\phi _{j})$ [recall
Eq.~(\ref{eq:Uij})]. Suppose first that we turn on this logic-gate
generating Hamiltonian in a manner that is neither very strong nor very
fast, so that the system-bath interaction is not negligible while $H_{S}$ is
on (this obviously puts less severe demands on experimental implementation).
Then the corresponding unitary operator describing the dynamics of system
plus bath is: 
\begin{equation}
\tilde{U}(t)=\exp [-it(H_{S}+H_{SB}+H_{B})].
\end{equation}
Now, \emph{if we choose }$H_{S}$\emph{\ so that it commutes with the
decoupling pulses}, then we can show that after decoupling 
\begin{equation}
\tilde{U}(t)\mapsto \exp [-i2t(H_{S}+H_{B})],  \label{eq:Ucomp}
\end{equation}
provided $t$ is sufficiently small. Tracing out the bath then leaves a
purely unitary, decoherence-free evolution on the system. To prove this,
assume we have chosen $t^{\prime }$ and the decoupling Hamiltonian $
H_{S}^{\prime }$ so that (i) $\exp (-it^{\prime }H_{S}^{\prime })H_{SB}\exp
(it^{\prime }H_{S}^{\prime })=-H_{SB}$, and (ii) $[H_{S}^{\prime },H_{S}]=0$
. Then 
\begin{eqnarray}
\tilde{U}(t)e^{-it^{\prime }H_{S}^{\prime }}\tilde{U}(t)e^{-it^{\prime
}H_{S}^{\prime }} &=&\tilde{U}(t)e^{-it[H_{S}+e^{-it^{\prime }H_{S}^{\prime
}}H_{SB}e^{it^{\prime }H_{S}^{\prime }}+H_{B}]} \\
&=&e^{-it(H_{S}+H_{SB}+H_{B})}e^{-it(H_{S}-H_{SB}+H_{B})} \\
&=&e^{-\{2it(H_{S}+H_{B})+t^{2}([H_{SB},H_{S}]+[H_{SB},H_{B}])+O(t^{3})\}},
\end{eqnarray}
where we have used the Baker-Campbell-Hausdorff formula, $\exp (\alpha
A)\exp (\alpha B)=\exp \{\alpha (A+B)+\frac{\alpha ^{2}}{2}[A,B]+O(\alpha
^{3})\}$.

Let us now show how to efficiently combine logic operations and decoupling
pulses. For simplicity consider only the case where we can neglect the $\bar{
X}$ error, i.e., our decoupling sequence is the 4-pulse one given in Eq.~(\ref{eq:4pulses}). Suppose we wish to implement a logical $X$ operation,
i.e., $\exp (-i\theta \overline{X}_{12})$. Recall that
this involves turning on the Hamiltonian $H_{S}^{X}=\Omega _{X}X_{\phi
}X_{\phi }\overset{\mathrm{DFS}}{\mapsto }\Omega _{X}\overline{X}_{12}$
between two physical qubits. Because the decoupling pulses $P=\exp (-i\frac{
\pi }{2}\overline{X}_{12})$ and $\Pi =\exp (\pm i\pi \overline{X}_{12})$ are
generated in terms of the same Hamiltonian, they commute with $H_{S}^{X}$
while eliminating $H_{SB}$ (except for the terms in $H_{SB}$ that have
trivial action on the DFS). Thus the conditions under which Eq.~(\ref
{eq:Ucomp}) were shown to hold are satisfied. This allows us to insert the
logic gates into the four free evolution periods involved in the pulse
sequence of Eq.~(\ref{eq:4pulses}). Thus, the full pulse sequence that
combines creation of collective dephasing conditions with execution of the
logic gate is: 
\begin{equation}
e^{-it(\Omega _{X}\overline{X}_{12}+H_{\mathrm{DFS}})}=\tilde{U}(t/4)\Pi 
\tilde{U}(t/4)P\tilde{U}(t/4)\Pi \tilde{U}(t/4)P^{\dagger },  \label{eq:X}
\end{equation}
with $\tilde{U}(t)=\exp [-it(H_{S}^{X}+H_{SB}+H_{B})]$, and which, using the
DFS encoding, is equivalent to the desired $\exp (-i\theta \overline{X}_{12})
$. This involves 8 control pulses, 4 of which are of the fast+strong type
(those involving $P$ and $\Pi $), and 4 of which must be fast, but need not
be so strong that we can neglect $H_{SB}$.

If we wish to implement logical $Y$ operation, i.e., $\exp (-i\theta 
\overline{Y}_{12})$, then we cannot now use $P$ and $\Pi $, since they
anticommute with $\overline{Y}_{12}$ and will eliminate it. Instead we
should use decoupling pulses generated in terms of $\overline{Y}_{12}$,
which will also have the desired effect of eliminating $H_{\mathrm{Leak}}$,
as well as $\bar{X}$ and $\bar{Z}$ logical errors, while commuting with the $
\overline{Y}$ logic operations (and for this reason can of course not
eliminate $\bar{Y}$ errors). These are just the $Q$ and $\Lambda $ pulses
defined in Eq.~(\ref{eq:PQetc}). In gate terms this implies turning on the Hamiltonian $H_{S}^{Y}=\Omega _{Y}X_{\phi
}X_{\phi +\pi /2}\overset{\mathrm{DFS}}{\mapsto }\Omega _{Y}\overline{Y}_{12}
$ between two physical qubits. Thus: 
\begin{equation}
e^{-it(\Omega _{Y}\overline{Y}_{12}+H_{\mathrm{DFS}})}=\tilde{U}(t/4)\Lambda 
\tilde{U}(t/4)Q\tilde{U}(t/4)\Lambda \tilde{U}(t/4)Q^{\dagger },
\label{eq:Y}
\end{equation}
with $\tilde{U}(t)=\exp [-it(H_{S}^{Y}+H_{SB}+H_{B})]$, and which, using the
DFS encoding, is equivalent to the desired $\exp (-i\theta \overline{Y}_{12})
$.

Finally, to generate single DFS-qubit rotations about an arbitrary axis we
can combine Eqs.~(\ref{eq:X}),(\ref{eq:Y}) according to the Euler angles
construction.\cite{Nielsen:book} Given that Eqs.~(\ref{eq:X}),(\ref{eq:Y})
each take 8 pulses, the Euler angle method will generate an arbitrary
DFS-qubit rotation in at most 24 pulses.

Concerning gates that entangle two DFS-qubits, the situation may be more
involved, since now the next-nearest neighbor pulses in Eq.~(\ref{eq:create4}), that create the collective dephasing conditions on four qubits, do not
all commute with the $U_{4}$ gate of Eq.~(\ref{eq:U4}). Therefore here we
must resort to the strong + fast method of the previous subsection, i.e., we
need to synchronize the $U_{4}$ pulses with the end of the decoupling pulse
sequence. We do not here analyze the situation with respect to the exchange
Hamiltonian implementation of conditional logic gates.

Taken together, the methods described in this section provide an explicit
way to implement universal QC in a manner that offers protection against all
sources of qubit decoherence, using a fast + strong (or fast + weak)
versions of logic gates.

\section{Discussion and Conclusions}

\label{conclusions}

We have proposed a method of encoded recoupling and decoupling (ERD) for
performing decoherence-protected quantum computation. Our method combines
an
encoding into qubit-pair decoherence-free subspaces (each pair yielding one
encoded qubit), with sequences of recoupling and decoupling pulses. The qubit
encoding protects against collective dephasing processes, while the
decoupling pulses symmetrize all other sources of decoherence into a
collective dephasing interaction. The recoupling pulses are used to
implement encoded quantum logic gates, either during or in between the
decoupling pulses.

The dynamical decoupling method
requires an exponential number of pulses if the most general form of
decoherence is to be suppressed, that can couple arbitrary numbers of qubits
to the environment (total decoherence \cite{Lidar:PRL98}). This exponential
scaling is avoided here by focusing on decoherence elimination inside blocks
of \emph{finite} size (e.g., at most four qubits) where arbitrary
decoherence is allowed. However, we have implicitly assumed that there are
no decoherence processes coupling different blocks. This is a reasonable
assumption for most quantum computer implementation, provided the different blocks can be kept
sufficiently far apart until they need to be brought together in order to
execute inter-block logic gates. When this happens, ERD can still be
efficiently applied on the temporarily larger block.

It may be questioned whether there is any advantage in using ERD compared to
methods of active quantum error correcting codes (QECC). Both ERD and QECC
are capable of dealing with arbitrary decoherence processes, and are fully
compatible with universal quantum computation. There are two main advantages
to ERD: First, we need only two physical qubits per logical qubit, compared to a redundancy
of five to one, in order to handle all single-qubit errors in QECC.\cite{Laflamme:96} So far the most advanced experiments outside NMR, i.e.,
those involving trapped ions, have used up to
four qubits,\cite{Sackett:00} so that this encoding economy is a distinct
advantage for near-term experiments. Second, our method is directly
compatible with Hamiltonians describing a variety of quantum computer proposals. On the other
hand it is not clear how to directly use QECC given
Hamiltonians describing specific systems. These are
general features of ERD: economy of encoding redundancy and use of only the
most easily controllable interactions. The disadvantage of ERD compared to
QECC is that there does not exist, at this point, a result analogous to the
threshold theorem of fault tolerant quantum error correction. This means
that we cannot yet guarantee full scalability of ERD as a stand-alone
method, because we do not yet know how to compensate for imperfect pulses.
However, in principle it is always possible to concatenate ERD with QECC, as
done, e.g., for DFS with QECC in Refs.~\citenum{KhodjastehLidar:02}, and then the standard
fault tolerance results apply. We expect
that the theory of composite pulses \cite{Cummins:02} will
also play a key role in this further development of ERD.

We hope that the methods proposed here will inspire
experimentalists to implement encoded recoupling and decoupling in the lab,
thus demonstrating the possibility of fully decoherence-protected quantum
computation.

\acknowledgments

This material is based on research sponsored by the Defense Advanced
Research Projects Agency under the QuIST program and managed by the Air
Force Research Laboratory (AFOSR), under agreement F49620-01-1-0468. D.A.L.
further gratefully acknowledges financial support from PRO, NSERC, and the
Connaught Fund.


\begin{thebibliography}{10}

\bibitem{Landauer:95}
R.~Landauer, ``Is quantum mechanics useful?,'' {\em Proc. Roy. Soc. London Ser.
  A} {\bf 353}, p.~367, 1995.

\bibitem{Unruh:95}
{W.G. Unruh}, ``{Maintaining coherence in quantum computers},'' {\em Phys. Rev.
  A} {\bf 51}, p.~992, 1995.

\bibitem{Shor:95}
{P.W. Shor}, ``{Scheme for reducing decoherence in quantum memory},'' {\em
  Phys. Rev. A} {\bf 52}, p.~2493, 1995.

\bibitem{Steane:96a}
{A.M. Steane}, ``{Error correcting codes in quantum theory},'' {\em Phys. Rev.
  Lett.} {\bf 77}, p.~793, 1996.

\bibitem{Kitaev:96}
{A.Yu. Kitaev}, ``{Quantum Computations: Algorithms and Error Corrections},''
  {\em Russian Math. Surveys} {\bf 52}, p.~1191, 1996.

\bibitem{Laflamme:96}
{R. Laflamme, C. Miquel, J.P. Paz and W.H. Zurek}, ``{Perfect Quantum Error
  Correction Code},'' {\em Phys. Rev. Lett.} {\bf 77}, p.~198, 1996.

\bibitem{Preskill:97a}
J.~Preskill, ``Reliable quantum computers,'' {\em Proc. Roy. Soc. London Ser.
  A} {\bf 454}, p.~385, 1998.

\bibitem{Knill:98}
{E. Knill, R. Laflamme and W. Zurek}, ``Resilient quantum computation,'' {\em
  Science} {\bf 279}, p.~342, 1998.

\bibitem{Steane:02}
{A.M. Steane}, ``{Overhead and noise threshold of fault-tolerant quantum error
  correction}.''
\newblock eprint quant-ph/0207119.

\bibitem{Gottesman:97a}
D.~Gottesman, ``Theory of fault-tolerant quantum computation,'' {\em Phys. Rev.
  A} {\bf 57}, p.~127, 1997.

\bibitem{Knill:97b}
{E. Knill and R. Laflamme}, ``{Theory of quantum error-correcting codes},''
  {\em Phys. Rev. A} {\bf 55}, p.~900, 1997.

\bibitem{WuLidar:01}
{L.-A. Wu and D.A. Lidar}, ``{Power of Anisotropic Exchange Interactions:
  Universality and Efficient Codes for Quantum Computing},'' {\em Phys. Rev. A}
  {\bf 65}, p.~042318, 2002.

\bibitem{LidarWu:01}
{D.A. Lidar and L.-A. Wu}, ``{Reducing Constraints on Quantum Computer Design
  by Encoded Selective Recoupling},'' {\em Phys. Rev. Lett.} {\bf 88},
  p.~017905, 2002.

\bibitem{WuLidar:01a}
{L.-A. Wu and D.A. Lidar}, ``{Qubits as Parafermions},'' {\em J. Math. Phys.}
  {\bf 43}, p.~4506, 2002.

\bibitem{WuLidar:01b}
{L.-A. Wu and D.A. Lidar}, ``{Creating Decoherence-Free Subspaces Using Strong
  and Fast Pulses},'' {\em Phys. Rev. Lett.} {\bf 88}, p.~207902, 2002.

\bibitem{ByrdLidar:01a}
{M.S. Byrd and D.A. Lidar}, ``{Comprehensive Encoding and Decoupling Solution
  to Problems of Decoherence and Design in Solid-State Quantum Computing},''
  {\em Phys. Rev. Lett.} {\bf 89}, p.~047901, 2002.

\bibitem{WuLidar:02}
{L.-A. Wu and D.A. Lidar}, ``{Universal Quantum Logic from Zeeman and
  Anisotropic Exchange Interactions},'' {\em Phys. Rev. A} {\bf 66}, p.~062314,
  2002.

\bibitem{WuByrdLidar:02}
{L.-A. Wu, M.S. Byrd, D.A. Lidar}, ``{Efficient Universal Leakage Elimination
  for Physical and Encoded Qubits},'' {\em Phys. Rev. Lett.} {\bf 89},
  p.~127901, 2002.

\bibitem{LidarWuBlais:02}
{D.A. Lidar, L.-A. Wu, A. Blais}, ``{Quantum Codes for Simplifying Design and
  Suppressing Decoherence in Superconducting Phase-Qubits},'' {\em Quant. Inf.
  Proc.} {\bf 1}, p.~155, 2002.

\bibitem{WuLidar:02a}
{L.-A. Wu, and D.A. Lidar}, ``{Universal Quantum Computation using Exchange
  Interactions and Teleportation of Single-Qubit Operations},''
\newblock eprint quant-ph/0208118.

\bibitem{LidarWu:02}
{D.A. Lidar and L.-A Wu}, ``{Encoded recoupling and decoupling: An alternative
  to quantum error correcting codes, applied to trapped ion quantum
  computation},'' {\em Phys. Rev. A., in press (2003)} .
\newblock eprint quant-ph/0211088.

\bibitem{Zanardi:97c}
{P. Zanardi and M. Rasetti}, ``{Noiseless Quantum Codes},'' {\em Phys. Rev.
  Lett.} {\bf 79}, p.~3306, 1997.

\bibitem{Duan:98}
{L.-M Duan and G.-C. Guo}, ``{Reducing decoherence in quantum-computer memory
  with all quantum bits coupling to the same environment},'' {\em Phys. Rev. A}
  {\bf 57}, p.~737, 1998.

\bibitem{Lidar:PRL98}
{D.A. Lidar, I.L. Chuang and K.B. Whaley}, ``{Decoherence free subspaces for
  quantum computation},'' {\em Phys. Rev. Lett.} {\bf 81}, p.~2594, 1998.

\bibitem{Lidar:00a}
{D.A. Lidar, D. Bacon, J. Kempe, and K.B. Whaley},
  ``\uppercase{D}ecoherence-free subspaces for multiple-qubit errors: (i)
  characterization,'' {\em Phys. Rev. A} {\bf 63}, p.~022306, 2001.

\bibitem{LidarWhaley:03}
{D.A. Lidar, K.B Whaley}, ``{Decoherence-Free Subspaces and Subsystems},''
  2003.
\newblock eprint quant-ph/0301032.

\bibitem{Viola:98}
L.~Viola and S.~Lloyd, ``Dynamical suppression of decoherence in two-state
  quantum systems,'' {\em Phys. Rev. A} {\bf 58}, p.~2733, 1998.

\bibitem{Duan:98e}
L.-M. Duan and G.~Guo, ``Suppressing environmental noise in quantum computation
  through pulse control,'' {\em Phys. Lett. A} {\bf 261}, p.~139, 1999.

\bibitem{Vitali:99}
D.~Vitali and P.~Tombesi, ``Using parity kicks for decoherence control,'' {\em
  Phys. Rev. A} {\bf 59}, p.~4178, 1999.

\bibitem{Vitali:01}
{D. Vitali and P. Tombesi}, ``{Heating and decoherence suppression using
  decoupling techniques},'' {\em Phys. Rev. A} {\bf 65}, p.~012305, 2002.

\bibitem{Zanardi:98b}
P.~Zanardi, ``Symmetrizing evolutions,'' {\em Phys. Lett. A} {\bf 258}, p.~77,
  1999.

\bibitem{Zanardi:99a}
P.~Zanardi, ``Computation on an error-avoiding quantum code and
  symmetrization,'' {\em Phys. Rev. A} {\bf 60}, p.~R729, 1999.

\bibitem{Zanardi:99d}
P.~Zanardi, ``Stabilizing quantum information,'' {\em Phys. Rev. A} {\bf 63},
  p.~012301, 2001.

\bibitem{Viola:99}
{L. Viola, E. Knill and S. Lloyd}, ``Dynamical decoupling of open quantum
  systems,'' {\em Phys. Rev. Lett.} {\bf 82}, p.~2417, 1999.

\bibitem{Viola:99a}
{L. Viola, E. Knill, and S. Lloyd}, ``{Universal Control of Decoupled Quantum
  Systems},'' {\em Phys. Rev. Lett.} {\bf 83}, p.~4888, 1999.

\bibitem{Viola:00a}
{L. Viola, E. Knill, and S. Lloyd}, ``{Dynamical Generation of Noiseless
  Quantum Subsystems},'' {\em Phys. Rev. Lett.} {\bf 85}, p.~3520, 2000.

\bibitem{Viola:01a}
{L. Viola}, ``{Quantum Control via Encoded Dynamical Decoupling},'' {\em Phys.
  Rev. A} {\bf 66}, p.~012307, 2002.

\bibitem{Viola:02}
{L. Viola and E. Knill}, ``{Robust dynamical decoupling with bounded
  controls},'' {\em Phys. Rev. Lett.} {\bf 90}, p.~037901, 2003.

\bibitem{Agarwal:01}
{G.S. Agarwal, M.O. Scully, and H. Walther}, ``{Inhibition of Decoherence due
  to Decay in a Continuum},'' {\em Phys. Rev. Lett.} {\bf 86}, p.~4271, 2001.

\bibitem{Search:00}
{C. Search and P.R. Berman}, ``{Suppression of Magnetic State Decoherence Using
  Ultrafast Optical Pulses},'' {\em Phys. Rev. Lett.} {\bf 85}, p.~2272, 2000.

\bibitem{ByrdLidar:01}
{M.S. Byrd and D.A. Lidar}, ``{Bang-Bang Operations from a Geometric
  Perspective},'' {\em Quant. Inf. Proc.} {\bf 1}, p.~19, 2001.

\bibitem{ByrdLidar:02}
{M.S. Byrd, D.A. Lidar}, ``{Empirical determination of dynamical decoupling
  operations},'' {\em Phys. Rev. A} {\bf 67}, p.~012324, 2003.

\bibitem{ByrdLidar:02a}
{M.S. Byrd and D.A. Lidar}, ``{Combined Error Correction Techniques for Quantum
  Computing Architectures},'' {\em J. Mod. Opt., in press (2003)} .
\newblock eprint quant-ph/0210072.

\bibitem{ShiokawaLidar:02}
{K. Shiokawa, D.A. Lidar}, ``{Dynamical Decoupling Using Slow Pulses: Efficient
  Suppression of $1/f$ Noise},'' 2002.
\newblock eprint quant-ph/0211081.

\bibitem{Uchiyama:02}
C.~Uchiyama and M.~Aihara, ``Multipulse control of decoherence,'' {\em Phys.
  Rev. A} {\bf 66}, p.~032313, 2002.

\bibitem{Bacon:99a}
{D. Bacon, J. Kempe, D.A. Lidar and K.B. Whaley}, ``{Universal Fault-Tolerant
  Computation on Decoherence-Free Subspaces},'' {\em Phys. Rev. Lett.} {\bf
  85}, p.~1758, 2000.

\bibitem{Kempe:00}
{J. Kempe, D. Bacon, D.A. Lidar, and K.B. Whaley}, ``{Theory of
  Decoherence-Free, Fault-Tolerant, Universal Quantum Computation},'' {\em
  Phys. Rev. A} {\bf 63}, p.~042307, 2001.

\bibitem{Lidar:CP01}
{D.A. Lidar, Z. Bihary, and K.B. Whaley}, ``{From Completely Positive Maps to
  the Quantum Markovian Semigroup Master Equation},'' {\em Chem. Phys.} {\bf
  268}, p.~35, 2001.

\bibitem{Kwiat:00}
{P.G. Kwiat, A.J. Berglund, J.B. Altepeter, and A.G. White}, ``{Experimental
  Verification of Decoherence-Free Subspaces},'' {\em Science} {\bf 290},
  p.~498, 2000.

\bibitem{Kielpinski:01}
{D. Kielpinski, V. Meyer, M.A. Rowe, C.A. Sackett, W.M. Itano, C. Monroe, and
  D.J. Wineland}, ``{A Decoherence-Free Quantum Memory Using Trapped Ions},''
  {\em Science} {\bf 291}, p.~1013, 2001.

\bibitem{Levy:01}
{J. Levy}, ``{Universal quantum computation with spin-1/2 pairs and Heisenberg
  exchange},'' {\em Phys. Rev. Lett.} {\bf 89}, p.~147902, 2002.

\bibitem{Benjamin:01}
{S.C. Benjamin}, ``{Simple Pulses for Universal Quantum Computation with a
  Heisenberg ABAB Chain},'' {\em Phys. Rev. A} {\bf 64}, p.~054303, 2001.

\bibitem{DiVincenzo:00a}
{D.P. DiVincenzo, D. Bacon, J. Kempe, G. Burkard, and K.B. Whaley},
  ``{Universal Quantum Computation with the Exchange Interaction},'' {\em
  Nature} {\bf 408}, p.~339, 2000.

\bibitem{Kempe:01}
{J. Kempe, D. Bacon, D.P. DiVincenzo and K.B. Whaley}, ``{Encoded Universality
  from a Single Physical Interaction},'' {\em Quant. Inf. Comp.} {\bf 1},
  p.~33, 2001.

\bibitem{Kempe:01a}
{J. Kempe and K.B. Whaley}, ``{Exact gate-sequences for universal quantum
  computation using the XY-interaction alone},'' {\em Phys. Rev. A} {\bf 65},
  p.~052330, 2001.

\bibitem{Vala:02}
{J. Vala and K.B. Whaley}, ``{Encoded Universality for Generalized Anisotropic
  Exchange Hamiltonians},'' {\em Phys. Rev. A} {\bf 66}, p.~022304, 2002.

\bibitem{Bacon:Sydney}
{D. Bacon, J. Kempe, D.P. DiVincenzo, D.A. Lidar, and K.B. Whaley}, ``Encoded
  universality in physical implementations of a quantum computer,'' in {\em
  Proceedings of the 1st International Conference on Experimental
  Implementations of Quantum Computation, Sydney, Australia},  R.~Clark, ed.,
  p.~257, Rinton, (Princeton, NJ), 2001.

\bibitem{Wineland:98}
{D. J. Wineland, C. Monroe, W. M. Itano, D. Leibfried, B. E. King, and D. M.
  Meekhof}, ``{Experimental Issues in Coherent Quantum-State Manipulation of
  Trapped Atomic Ions},'' {\em {J. of Res. of the National Inst. of Standards
  and Technology}} {\bf 103}, p.~259, 1998.
\newblock http://nvl.nist.gov/pub/nistpubs/jres/103/3/cnt103-3.htm.

\bibitem{Sorensen:00}
A.~S$\o$rensen and K.~M$\o$lmer, ``{Entanglement and quantum computation with
  ions in thermal motion},'' {\em Phys. Rev. A} {\bf 62}, p.~022311, 2000.

\bibitem{Sackett:00}
{C. A. Sackett, D. Kielpinski, B. E. King, C. Langer, V. Meyer, C. J. Myatt, M.
  Rowe, Q. A. Turchette, W. M. Itano, D. J. Wineland, and C. Monroe},
  ``{Experimental entanglement of four particles},'' {\em Nature} {\bf 404},
  p.~256, 2000.

\bibitem{Kielpinski:02}
{D. Kielpinski, C. Monroe and D.J. Wineland}, ``{Architecture for a large-scale
  ion-trap quantum computer},'' {\em Nature} {\bf 417}, p.~709, 2002.

\bibitem{Nielsen:book}
{M.A. Nielsen and I.L. Chuang}, {\em {Quantum Computation and Quantum
  Information}}, {Cambridge University Press}, Cambridge, UK, 2000.

\bibitem{Ernst:book}
{R.R. Ernst, G. Bodenhausen and A. Wokaun}, {\em {Principles of Nuclear
  Magnetic Resonance in One and Two Dimensions}}, no.~14 in {International
  Series of Monographs on Chemistry}, {Oxford University Press}, Oxford, 1997.

\bibitem{Palao:02}
{J.P. Palao and R. Kosloff}, ``{Quantum Computing by an Optimal Control
  Algorithm for Unitary Transformations},'' {\em Phys. Rev. Lett.} {\bf 89},
  p.~188301, 2002.

\bibitem{Tian:00}
{L. Tian and S. Lloyd}, ``{Resonant cancellation of off-resonant effects in a
  multilevel qubit},'' {\em Phys. Rev. A} {\bf 62}, p.~050301, 2000.

\bibitem{KhodjastehLidar:02}
{K. Khodjasteh and D.A. Lidar}, ``{Universal Fault-Tolerant Quantum Computation
  in the Presence of Spontaneous Emission and Collective Dephasing},'' {\em
  Phys. Rev. Lett.} {\bf 89}, p.~197904, 2002.

\bibitem{Cummins:02}
{H.K. Cummins, G. Llewellyn, J.A. Jones}, ``{Tackling Systematic Errors in
  Quantum Logic Gates with Composite Rotations},''
\newblock eprint quant-ph/0208092.

\end{thebibliography}

\end{document}